\RequirePackage{fix-cm}

\documentclass[twocolumn]{svjour3}          
\smartqed  

\usepackage{url}
\usepackage{graphicx}
\usepackage{subfig}
\usepackage{amsmath}
\usepackage{mathtools}
\usepackage{array}
\usepackage{natbib}
\usepackage{multirow}
\usepackage{rotating}
\usepackage{breqn}

\journalname{Bulletin of Earthquake Engineering}
\begin{document}

\title{Developing and Testing the Automated Post-Event Earthquake Loss Estimation and Visualisation (APE-ELEV) Technique}

\titlerunning{Automated Earthquake Loss Estimation and Visualisation}     

\author{Anthony Astoul, Christopher Filliter, Eric Mason, Andrew Rau-Chaplin, Kunal Shridhar, Blesson Varghese and Naman Varshney}

\authorrunning{Astoul, Filliter, Mason, Rau-Chaplin, Shridhar, Varghese and Varshney}

\institute{\emph{B. Varghese} (Corresponding Author) \at Big Data Lab, Faculty of Computer Science\\University of St Andrews, Scotland, UK\\E-mail: varghese@st-andrews.ac.uk\\URL: http://www.blessonv.com
\and
A. Astoul, C. Filliter, E. Mason, K. Shridhar, A. Rau-Chaplin and N. Varshney \at Risk Analytics Lab, Faculty of Computer Science\\Dalhousie University, Halifax, Nova Scotia, Canada
}

\date{Received: 5 September 2012 / Accepted: 22 June 2013}

\maketitle

\begin{abstract}
An automated, real-time, multiple sensor data source relying and globally applicable earthquake loss model and visualiser is desirable for post-event earthquake analysis. To achieve this there is a need to support rapid data ingestion, loss estimation and integration of data from multiple data sources and rapid visualisation at multiple geographic levels. In this paper, the design and development of the Automated Post-Event Earthquake Loss Estimation and Visualisation (APE-ELEV) system for real-time estimation and visualisation of insured losses incurred due to earthquakes is presented. A model for estimating ground up and net of facultative losses due to earthquakes in near real-time is implemented. Since post-event data is often available immediately from multiple disparate sources, a geo-browser is employed to facilitate the visualisation and integration of earthquake hazard, exposure and loss data. The feasibility of APE-ELEV is demonstrated using a test case earthquake that occurred in Tohoku, Japan (2011). The APE-ELEV model is further validated for ten global earthquakes using industry loss data. 
\end{abstract}

\section{Introduction}
\label{intro}
Research in estimating losses for catastrophes have led to the development of a wide variety of earthquake loss models. Earthquake loss models can generate loss values before an event occurs or while an event is evolving or after an event occurs. Earthquake loss models can be classified as probabilistic, deterministic and real-time models. Probabilistic models produce a maximum probable loss value using a stochastic event catalog which represents a sample of possible future earthquakes. Models such as CAPRA - Central American Probabilistic Risk Assessment \citep{CAPRA}, EQRM - Earthquake Risk Model \citep{EQRM} and RiskScape \citep{RiskScape} are probabilistic models. In deterministic models the losses caused by a specific event that occurred are estimated. LNECLOSS \citep{LNECLOSS}, REDARS - Risks from Earthquake Damage to Roadway Systems \citep{REDARS} and NHEMATIS \citep{NHEMATIS} are deterministic models. Real-time models estimate losses soon after (near real-time) an earthquake has occurred. Examples include ELER - Earthquake Loss Estimation Routine \citep{ELER}, EmerGeo \citep{EmerGeo} and PAGER - Prompt Assessment of Global Earthquakes for Response \citep{PAGER}. A hybrid of the former models are seen in HAZUS (combines deterministic, probabilistic and real-time models) \citep{HAZUS}, KOERILOSS \citep{KOERILOSS} and MAEviz \citep{MAEviz}. In this paper, a loss estimator which produces loss values in near real-time and can model past earthquake events is presented. 

Models that focus on generating a probable loss value use a catalog of possible future earthquakes. In such models, there is no focus on a specific event and any analysis is done before an earthquake may occur and is called pre-event analysis. Examples include AIR \citep{AIR}, DBELA - Displacement-Based Earthquake Loss Assessment \citep{DBELA} and MDLA \citep{MDLA}. For quick and imminent decision making it is desirable that loss estimates be accurately generated as an event evolves. Post-event analysis presents a timely evaluation of losses due to an earthquake in the minutes, hours, days and weeks immediately following an earthquake. Examples of post-event models are INLET - Internet-based Loss Estimation Tool \citep{INLET}, PAGER \citep{PAGER} and Extremum \citep{Extremum}. Models combining both pre-event and post-event analysis are available in EPEDAT - Early Post-Earthquake Damage Assessment Tool \citep{EPEDAT}, HAZUS - \citep{HAZUS} and SELENA - SEismic Loss Estimation using a logic tree Approach \citep{SELENA}. The model proposed in this paper focuses on analysing the effects of an earthquake soon after it occurs and modelling the effects of a past earthquake. 

Pre-event models are of limited interest in the context of estimating losses in real-time. In this paper the focus is on post-event analysis since it is different from pre-event analysis in a number of important ways:
\begin{itemize}
	\item[(a)] the focus is on a single earthquake event which has just occurred rather than a catalog of possible future events, or on a past earthquake event which can be modelled from archived sensory data.
	\item[(b)] there is an evolving view of the event as it unfolds, and therefore the sensor data related to the event changes hours, days and weeks after the event,
	\item[(c)] there is a need for rapid estimation of losses to guide early responses \citep{EarlyWarning}, and
	\item[(d)] since post-event data is available from multiple sources, there is a need to visualise and integrate hazard, exposure and loss data from these multiple sources.
\end{itemize}

The 2011 Tohoku earthquake that struck off the Pacific coast of Japan at 05:46 UTC on Friday, 11 March 2011 is a recent example that illustrates the importance of post-event analysis. Figure 1 presents the timeline of the earthquake. Fifteen alerts $A_{1}-A_{15}$ were issued by PAGER/ShakeMap in time periods ranging from within an hour to six months after the earthquake. The first alert was issued twenty three minutes after the event and reported a magnitude 7.9 earthquake. Additional information such as initial Peak Ground Velocity and Peak Ground Acceleration maps of the ground shake was also available with the alert. Further, over the course of the first day alone four additional alerts were issued each updating the data available. Not only did the earthquake event unfold over time but the data describing the event and our knowledge of the event evolved. The earthquake data alone was not sufficient to produce reliable loss estimates because between 06:15 UTC and 07:52 UTC a tsunami struck the coastal towns. Additional data sources are required for complete loss estimation.

\begin{figure*}
	\centering
	\includegraphics[width = \textwidth]{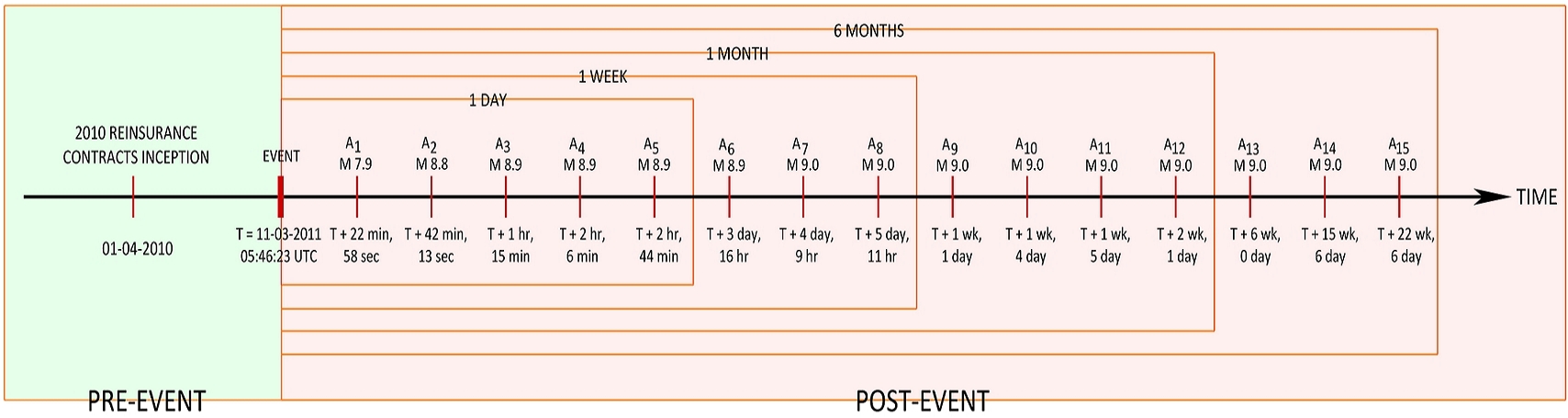}
	\caption{Timeline of the 2011 Tohoku Earthquake}
	\label{figure1}
\end{figure*}

Estimating loss values of a future earthquake is based on using a static catalog containing data related to historic events and is employed in pre-event analysis. For example, models such as AIR \citep{AIR}, DBELA \citep{DBELA} and EQRM \citep{EQRM} employ static catalogs. A static catalog therefore is not sufficient to estimate accurate losses as an earthquake evolves over hours and days of its occurrence. There is a need for up-to-date information of an earthquake as it evolves. One possibility is to make use of seismic sensor networks which can provide earthquake information as soon as minutes after it has occurred. Shakemaps \citep{5,6}, for example, are a representation of earthquake sensory information. Models that employ real-time models include EmerGeo \citep{EmerGeo}, INLET \citep{INLET} and PAGER \citep{PAGER}. A few models incorporate both historic and sensor data such as in HAZUS \citep{HAZUS}, MDLA \citep{MDLA} and SELENA \citep{SELENA}. In this paper, we investigate how sensor data from multiple sources can be used for timely estimation of losses. 

The use of regional seismic sensor networks can provide a model with only region specific data and thereby restricts loss estimation to regions. This may be due to the nature of the research where the project was undertaken and therefore only a country or a region was considered. Models such as OpenRisk \citep{OpenRisk}, TEFER - Turkish Emergency Flood and Earthquake Recovery Programme Earthquake Model \citep{TEFER} and TELES - Taiwan Earthquake Loss Estimation System \citep{TELES} are examples that analyse earthquakes in a region. To ensure global applicability of the model it needs to rely on global sensor networks. EPEDAT \citep{EPEDAT}, RADIUS \citep{RADIUS} and QLARM - Earthquake Loss Assessment for Response and Mitigation \citep{QLARM} are a few examples. Further, full-fledged global applicability also implies being able to use the model to estimate losses at different geographic levels (for example, loss estimation at cities, counties, states and countries). The model presented in this paper explores how global applicability can be achieved.

Among the earthquake loss estimation models that have been referenced, ELER, EmerGeo, EPEDAT, Extremum, HAZUS, INLET, PAGER, QLARM, QUAKE-LOSS, SELENA and TELES support post-event analysis. Among these, models such as, ELER, EPEDAT, HAZUS, INLET and TELES are region restricted. While these models may provide close to accurate loss estimates, yet they do not support global earthquakes. This may be due to the reliance of the models on regional seismic networks. 

The EmerGeo earthquake model produces maps of MMI and Peak Ground Acceleration (PGA) and can predict damages. Loss estimates are not a focus in the model. Both the Extremum and QUAKELOSS models rely on multiple data sources but are focused on structural and human losses. Financial loss estimates are not considered in both models. PAGER (Prompt Assessment of Global Earthquakes for Response) provides fatality and economic loss impact estimates. However, PAGER does not determine region specific loss data. Global financial and economic organisations need to know the losses (estimates) incurred at different geographical levels. The QLARM model calculates human losses and damage in a given human settlement. However, QLARM does not focus on estimating financial losses. The SELENA model and the complementing RISe (Risk Illustrator for SELENA) \citep{RISe} visualisation software computes real-time loss estimates and presents the losses visually. However, there seems to be less automation along the pipeline from obtaining real-time data to visualising the losses. The real-time data needs to be provided by the user to the SELENA model. Research that is pursued for automated post-event estimation of financial losses globally is sparse at best, though many loss models are available in the public domain \citep{OPAL}.

The research reported in this paper is motivated towards the development of (a) a real-time, (b) a post-event, (c) a multiple sensor data relying and (d) a globally applicable loss model. To achieve this there is a need to support rapid data ingestion, rapid loss estimation, rapid visualisation and integration of data from multiple data sources and rapid visualisation at multiple geographic levels. 

The \textbf{A}utomated \textbf{P}ost-\textbf{E}vent \textbf{E}arthquake \textbf{L}oss \textbf{E}stimation and \textbf{V}isualisation (APE-ELEV) system is proposed, which comprises three primary modules, namely the Earthquake Loss Estimator (ELE), the Earthquake Visualiser (EV) and the ELEV Database (ELEV-DB). The ELE module is built on PAGER and Shakemap for accessing real-time earthquake data and estimating losses at different geographic levels. The ELE module computes financial losses. Visualisation of the losses is facilitated by the EV module. The ELEV-DB module aids the functioning of the ELE and EV modules.

The remainder of this paper is organised as follows. Section \ref{centralisedarchitecture} proposes a centralised architecture for the Automated Post-Event Earthquake Loss Estimator and Visualiser (APE-ELEV). The loss estimation module is presented in Section \ref{estimator} and the loss visualiser module is presented in Section \ref{visualiser}. Section \ref{distributedarchitecture} presents a distributed architecture for the APE-ELEV and how estimation and visualisation are distributed across the server and the client respectively. Section \ref{experiments} presents one test case using APE-ELEV and a validation study of the model using ten global earthquakes. Section \ref{conclusion} concludes the paper. 

\section{Centralised APE-ELEV Architecture}
\label{centralisedarchitecture}
The Automated Post-Event Earthquake Loss Estimator and Visualiser (APE-ELEV) is a system that determines expected losses due to the occurrence of an earthquake (on building that are exposed to the earthquake, otherwise called exposure) and graphically display these losses. Decision makers in financial organisations, governmental agencies working toward disaster management and emergency response teams can benefit from interpreting the output produced by APE-ELEV for aiding imminent decision making. The output can also be adjusted for the benefit of the decision maker by changing the exposure data. 

The APE-ELEV system determines two types of losses. Firstly, the Ground Up Loss, referred to as GUL which is the entire amount of an insurance loss, including deductibles, before applying any retention or reinsurance. Secondly, the Net of Facultative Loss, referred to as NFL which is the entire amount of an insurance loss, including deductibles, primary retention and any reinsurance. The determined losses can be visualised at four geographic levels, namely country, state, county and city, on a geo-browser. The country, state and county levels are sometimes referred to as regions, while the city level is referred to as both point and population centre. Indicators are defined to facilitate visualisation at the region level; indicators are either event-specific (for example, losses at regions) or geography-specific (for example, population at cities or regions).

APE-ELEV is composed of three primary modules, namely the Earthquake Loss Estimator and Visualiser Database (ELEV-DB), the Earthquake Loss Estimator (ELE) and the Earthquake Visualiser (EV). Figure 2 shows the architecture of APE-ELEV. The ELEV-DB module is a collection of tables related to an event and geographic data. The ELE model (see Figure 2 (top)) as the name suggests estimates the losses incurred when an earthquake occurs. The EV model (see Figure 2 (bottom)) again as the name suggests facilitates the visualisation of the loss estimates generated by the ELE model.

\begin{figure*}
	\centering
	\includegraphics[width = 0.8\textwidth]{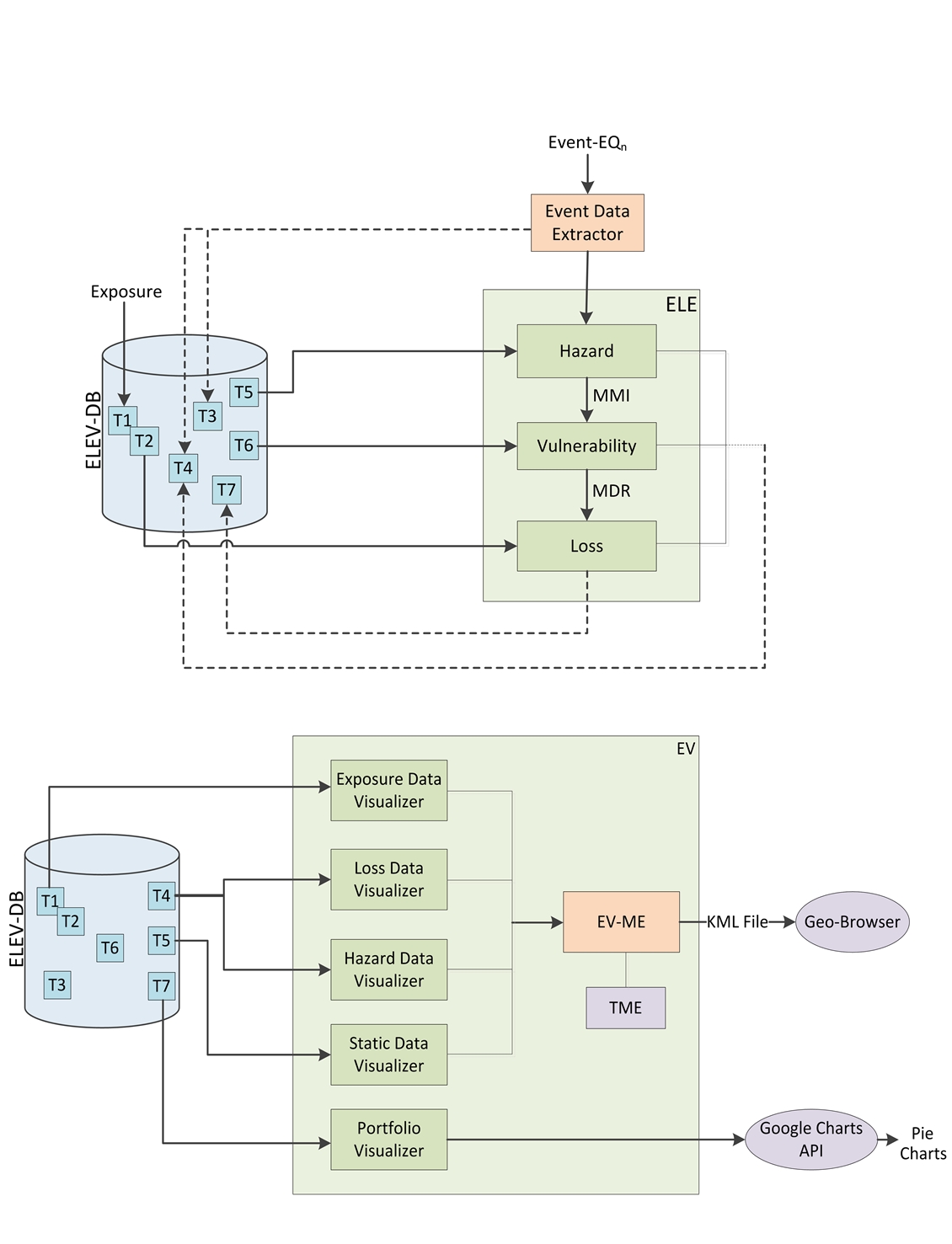}
	\caption{The APE-ELEV architecture comprising the ELE (top), EV (bottom) and ELEV-DB modules. Legend as follows - {T1}: Ground Up Exposure; T2: Net of Facultative Exposure; T3: Event Data; T4: Indicator Values; T5: Geographical Information; T6: MDR Data; T7: Loss Data. MMI: Modified Mercalli Intensity; MDR: Mean Damage Ratio; EV: Earthquake Visualiser; ELE: Earthquake Loss Estimator; TME: Thematic Mapping Engine; EV-ME: Earthquake Visualiser Mapping Engine; ELEV-DB: Earthquake Loss Estimator and Visualiser Database}
	\label{figure2}
\end{figure*}

The ELEV-DB module comprises seven tables which contribute to the working of the ELE and the EV modules. The tables are:
\begin{enumerate}
\item[(i)] $T_{1}$, which consists of industrial data for Ground Up Exposure, 
\item[(ii)] $T_{2}$, which consists of industrial data for Net of Facultative Exposure, 
\item[(iii)] $T_{3}$, which consists of event data, 
\item[(iv)] $T_{4}$, which consists of a set of indicators, 
\item[(v)] $T_{5}$, which consists of geographic information that is used to map lower geographic levels onto higher geographic levels (for example, mapping of cities onto counties or counties onto state), 
\item[(vi)] $T_{6}$, which consists of data that is generated from the Jaiswal and Wald Mean Damage Ratio (MDR) model \citep{14}, and 
\item[(vii)] $T_{7}$, which comprises loss data populated by the ELE module.
\end{enumerate}

The ELE module, as shown in Figure 2 (top), comprises three sub-modules, namely the Hazard, Vulnerability and Loss modules. The Hazard module receives two inputs, firstly, the data on cities (i.e. population centres with more than one thousand people) affected by the earthquake, and secondly, geographic information required for mapping lower geographic levels onto higher geographic levels. The Hazard module produces the measure of severity of an earthquake, otherwise referred to as the Modified Mercalli Intensity (MMI), in a city and region. The MMI values along with data from $T_{6}$ are used by the Vulnerability module to produce MDR values. This data is employed by the Loss module along with two types of exposure data, namely Ground Up Exposure and Net of Facultative Exposure to generate both the GUL and NFL losses. The Event Data Extractor receives the notification of the event and initiates the ELE.

The EV module, as shown in Figure 2 (bottom), comprises five sub-modules, namely the Exposure Data Visualiser, Loss Data Visualiser, Hazard Data Visualiser, Static Data Visualiser and the Portfolio Visualiser. The visualiser modules employ a geo-browser for graphical display. The Exposure Data Visualiser presents the exposure for different geographic levels. The Loss Data Visualiser presents the GUL and NFL for different geographic levels. The Hazard Data Visualiser presents the MMI and MDR for different geographic levels. Static Data Visualiser is employed for presenting geography-specific indicators, and as the name implies these indicator values do not change from one event to another. The Portfolio Visualiser presents a comparison of losses and exposures. The Earthquake Visualiser Mapping Engine (EV-ME) module facilitates visualisation of data on a geo-browser.

Having presented the architecture of APE-ELEV, it is also necessary to consider how the ELE, EV and ELEV-DB modules and their sub-modules glue together for coherent functioning. The data required to kick-start APE-ELEV is obtained before the occurrence of an earthquake or in a pre-event phase. An Accumulation Model is used to generate the Ground Up and Net of Facultative exposures at the region level. Casualties are proportional to the number of people present in the affected area and the quantity and value of buildings, infrastructure and other property in this area. The Accumulation Model quantifies regional exposure based on the whether economic losses need to be determined for the assets insured by the insurance/reinsurance company. In the research reported in this paper, the Accumulation Model is a black box used by the industrial partner supporting this research and the model generated GUL and NFL exposures for a given region. The region level exposure is then disaggregated into cities (i.e., population centres that fall within the region) based on the percentage of population. The city level exposure is further used by the ELE module in the post-event phase.

\section{The ELE Module}
\label{estimator}
For an earthquake event, $EQ_{n}$, that has just occurred or is unfolding we firstly need to be notified of the event. An automated system for notifying earthquakes is ShakeCast Lite \cite{15}. The ELE module employs ShakeCast Lite for notification alerts which are received by the Event Data Extractor. When the notification alert is received the ELE module is instantiated. Further, we require real-time data of the earthquake. The Prompt Assessment of Global Earthquakes for Response (PAGER) is an automated system that can provide such real-time data. The ELE module employs the real-time data from PAGER/Shakemap that is acquired as an .xml file. The .xml file is then parsed to extract event related information that is stored in $T_{3}$ of ELEV-DB. Information such as an affected city, represented as $L_1$ ($L_1$ represents city, $L_2$ represents counties, and $L_3$ represents states and $L_4$ represents countries), population of the city, represented as $P(L_1)$ and MMI of the city, represented as $MMI(L_1)$ is provided to the hazard module.

The hazard module computes the MMI at higher geographic levels using the MMI of affected cities. If the geographic level is represented as $L_n$, where $n = 2, 3$ and $4$, the population at the geographic level $L_n$ is represented as $P(L_n)$ and the MMI at the geographic level $L_n$ is represented as $MMI(L_n)$, then
\begin{equation}
\label{equation1}
MMI({{L_{(n)}}_i}) = \frac{\sum\limits_{j=1}^{q}MMI\big({{L_{(n-1)}}_j}\big) \times P\big({L_{(n-1)}}_j\big)}{\sum\limits^{q}_{j=1}P\big({L_{(n-1)}}_{j}\big)}\\
\end{equation}

\noindent where $i=1, 2, \cdots p$ ($p$ is the total no of affected regions), and $j=1,2, \cdots q$ ($q$ is the number of affected cities in a region $i$). The geographic data to evaluate whether an affected city lies within a given region is provided through $T_{5}$.

The double subscript notation is used to capture the idea that there are population centres which are affected due to the earthquake within a large affected region. For example, consider an earthquake that affects two counties, $county_{1}$ and $county_{2}$. In the equation counties are represented by $L_{2}$ and since there are two affected counties, $p = 2$, and $i$ iterates two times. 

Assume there are three cities in $county_{1}$, namely $city_{1}$, $city_{2}$ and $city_{3}$, their populations denoted as $P(city_{1})$, $P(city_{2})$ and $P(city_{3})$ and their MMIs denoted as $MMI(city_{1})$, $MMI(city_{2})$ and $MMI(city_{3})$ respectively. For this county $q = 3$ (three cities are in the affected region, and $j$ iterates three times for this county).

The MMI at the county levels $MMI(L_{2})$ for $county_{1}$ is equal to 
\begin{eqnarray*}
MMI(county_{1})	& = 	& \Big[\Big(MMI(city_{1}) \times P(city_{1})\Big) +\\
		& 	& \Big(MMI(city_{2}) \times P(city_{2})\Big) +\\
		&	& \frac{\Big(MMI(city_{3}) \times P(city_{3})\Big)\Big]}{P(city_{1}) + P(city_{2}) + P(city_{3})}
\end{eqnarray*}

Assume four cities in $county_{2}$, namely $city_{4}$, $city_{5}$, $city_{6}$ and $city_{7}$, their populations denoted as $P(city_{4})$, $P(city_{5})$, $P(city_{6})$ and $P(city_{7})$ and their MMIs denoted as $MMI(city_{4})$, $MMI(city_{5})$, $MMI(city_{6})$ and $MMI(city_{7})$ respectively. For this county $q = 4$ (four cities are in the affected region, and $j$ iterates four times for this county). 

The MMI at the county levels $MMI(L_{2})$ for $county_{2}$ is
\begin{eqnarray*}
MMI(county_{2})	& = 	& \Big[\Big(MMI(city_{4}) \times P(city_{4})\Big) + \\
		&	& \Big(MMI(city_{5}) \times P(city_{5})\Big) + \\
		&	& \Big(MMI(city_{6}) \times P(city_{6})\Big) + \\
		&	& \frac{\Big(MMI(city_{7}) \times P(city_{7})\Big)\Big]}{\Big(P(city_{4}) + P(city_{5}) + } \\
		&	& P(city_{6}) + P(city_{7})\Big)
\end{eqnarray*}

Consider that both counties, $county_{1}$ and $county{2}$, are in the same state, $state_{1}$, the population of the counties denoted as  $P(county_{1})$ and $P(county_{2})$, and the MMIs of the counties obtained from the above equations. 

The MMI at the state level $MMI(L_{3})$ for $state_{1}$ is
\begin{eqnarray*}
MMI(state_{1}) 	& = & \Big[\Big(MMI(county_{1}) \times P(county_{1})\Big) + \\
				&	&\frac{\Big(MMI(county_{2}) \times P(county_{2})\Big)\Big]}{P(county_{1}) + P(county_{2})}
\end{eqnarray*}

The $MMI(L_n)$, where $n = 1, 2, 3$ and $4$ is then utilised by the Vulnerability module to compute $MDR(L_n)$. Unlike the Hazard module, the city level is considered in the Vulnerability module, and therefore $n$ ranges from 1 to 4. It is worthwhile to note that MMI values range from I to XII. $T_{6}$ which was originally generated by the Jaiswal and Wald MDR model provides the MDR value corresponding to an integer MMI value. Should a floating point MMI value be obtained during computations from the hazard module, then the MDR values are computed by linear interpolation in the Vulnerability module. For example, if MMI is obtained as 7.5 from the Hazard module, then the MDR values corresponding to MMI VII and MMI VIII are interpolated in the Vulnerability module to obtain the MDR value for MMI-7.5. Such a technique is employed in HAZUS \citep{HAZUS}.

The MDR value of a city is provided to the Loss module, along with the Ground Up and the Net of Facultative exposure data from $T_{1}$ and $T_{2}$. The GUL and NFL of a city are computed by multiplying the MDR values for a city with the exposure of the city. The city losses are then aggregated onto higher geographic levels using $T_{5}$ to compute the losses on the county, state and country levels. The total loss corresponding to an event is provided to $T_{3}$, while the regional losses corresponding to an event is provided to $T_{4}$ and losses related to a specific line of business in $T_{7}$. Line of business refers to a statutory set of insurance/reinsurance policies to define coverage. The coverage may or may not affect a strategic business unit. The hierarchies structures of lines of business are property - fire insurance, business interruption and natural catastrophes; casualty - liability, motor, non-life accident and health; special lines - aviation, engineering, marine; credit and surety. These lines of business are either industrial, personal or commercial coverages. 

The ELEV-DB module plays an important role in providing data to and receiving data from the ELE module. During the period from the notification of an event until completion of computing losses, tables $T_{3}$, $T_{4}$ and $T_{7}$ are modified. Tables $T_{1}$, $T_{2}$, $T_{5}$ and $T_{6}$ provide input to the ELE module.

\section{The EV Module}
\label{visualiser}
The five sub-modules of EV, namely the Exposure Data Visualiser, the Loss Data Visualiser, the Hazard Data Visualiser, the Static Data Visualiser and the Portfolio Visualiser operate in parallel. This is unlike the ELE sub-modules that operate in sequence. The functioning of the sub-modules of EV are nevertheless presented sequentially in this section for the sake of convenience.

The Exposure Data Visualiser utilises $T_{1}$ and $T_{2}$ for displaying two types of exposures, the Ground Up Exposure and the Net of Facultative Exposure. The latitude, longitude and geography related indicators of all regions are extracted from $T_{5}$ and provided to the Earthquake Visualiser Mapping Engine (EV-ME). The EV-ME module generates a .kml (Keyhole Markup Language) file that contains place marks which highlight the exposure of the regions. The .kml format is compatible for visualisation on Geo-browsers \citep{16}, and in this research Google Earth is employed. The Thematic Mapping Engine (TME) is the underlying building block of EV-ME \citep{17}. A number of visualisation techniques such as bar, prism, choropleth, collada and push pins are made available for facilitating analysis of the data.

The Loss Data Visualiser utilises $T_{4}$ from which regional loss data is extracted for displaying the Ground Up and Net of Facultative losses. Similar to the Exposure Data Visualiser, the EV-ME module generates a .kml file that is viewable on Google Earth.

The Hazard Data Visualiser utilises $T_{4}$ and $T_{5}$ from which regional and point hazard data are extracted respectively for displaying MMI and MDR at all geographic levels. Similar to the above modules a .kml file is generated by the EV-ME module.

The Static Data Visualiser again utilises $T_{4}$ and $T_{5}$ from which cities affected by the event and static-data related to the affected cities are extracted respectively. A .kml file is generated by the EV-ME module and the extracted data is visualised.

The Portfolio Visualiser that is incorporated within the EV module compares losses and exposure (of areas affected by the event) by line of business. Data related to the distribution of total losses by line of business such as industrial, personal and commercial is extracted from $T_{7}$. Since visualisations are provided on pie-charts, the EV-ME module is not employed.

\section{Distributed APE-ELEV Architecture}
\label{distributedarchitecture}

The distributed APE-ELEV comprises the server system and the client system, as shown in Figure 3, and are considered in the following sub-sections.

\begin{figure*}
	\centering
	\includegraphics[width = 0.9\textwidth]{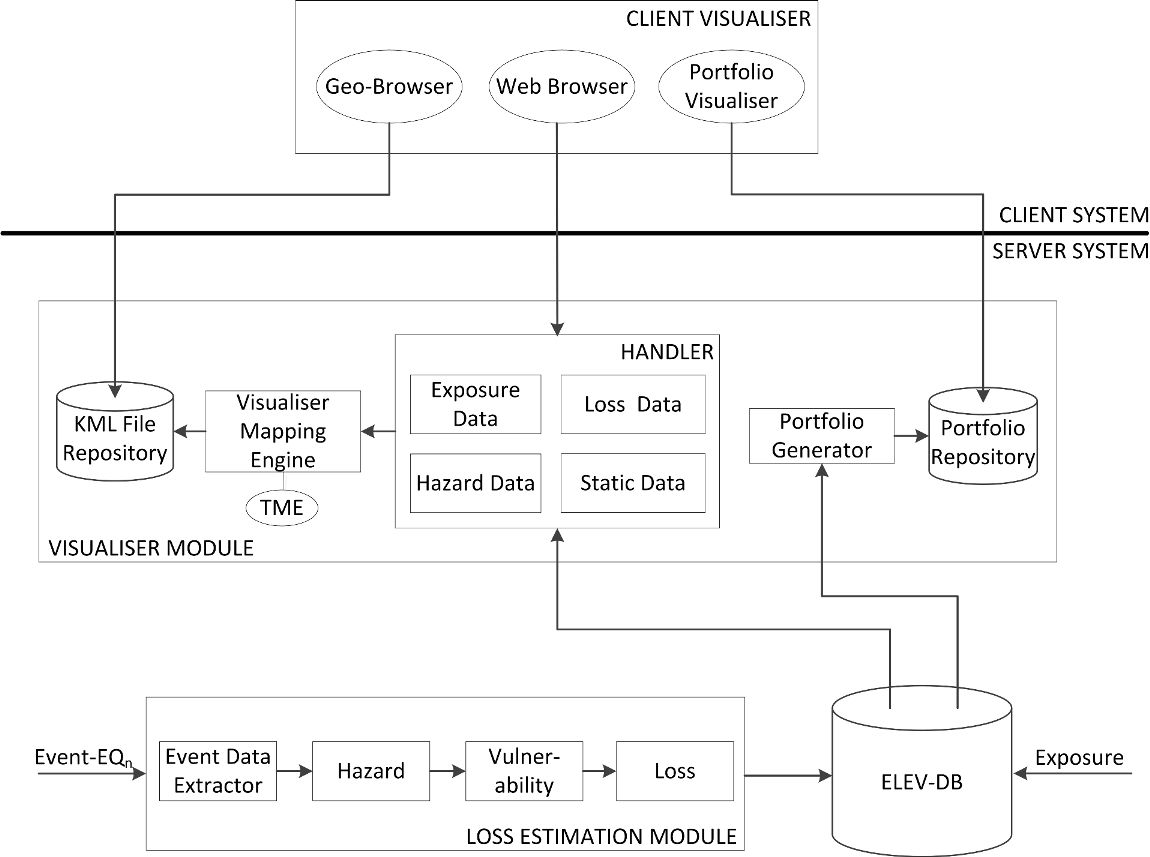}
	\caption{The Distributed APE-ELEV architecture}
	\label{figure3}
\end{figure*}

\subsection{Server and Client System}
The APE-ELEV server system consists of the ELEV-DB database, the ELE module and an EV module. The ELEV-DB and the ELE module are similar to those employed in the centralised architecture. The EV module is different from the centralised architecture as the geo-browser, the web browser and the portfolio visualiser are located on the client system. 

To facilitate the handling of client requests, an additional sub-module is required on the server visualiser system, and therefore the data handler is employed which acts as an interface between client requests and the data available for visualisation that is stored in the database. Four handlers are available, namely the exposure data handler, the hazard data handler, the loss data handler and the static data handler. The exposure data handler retrieves the exposure for different geographic levels. The loss data handler retrieves GUL and NFL for different geographic levels. The hazard data handler retrieves MMI and MDR for different geographic levels. The static data handler retrieves geography-specific indicators. 

The mapping engine receives data from the handlers and facilitates the visualisation of data on the client system. It is built on the Thematic Mapping Engine (TME) \citep{17} and generates .kml files. The KML file repository stores the .kml files generated by the mapping engine. The portfolio generator is built on the Google Chart API and presents a comparison of losses and exposures as pie-charts. 

The client system in the distributed APE-ELEV is a Client Visualiser that consists a geo-browser, an event viewer and a portfolio viewer. 

\subsection{Communication Sequence between the Client-Server modules}

Figure 4 is the illustration of interactions between the client and server modules. The loss estimation module executes Step 1 to Step 5 after it receives an earthquake notification, thereby storing loss values in the database. 

\begin{figure*}
	\centering
	\includegraphics[width = 0.8\textwidth]{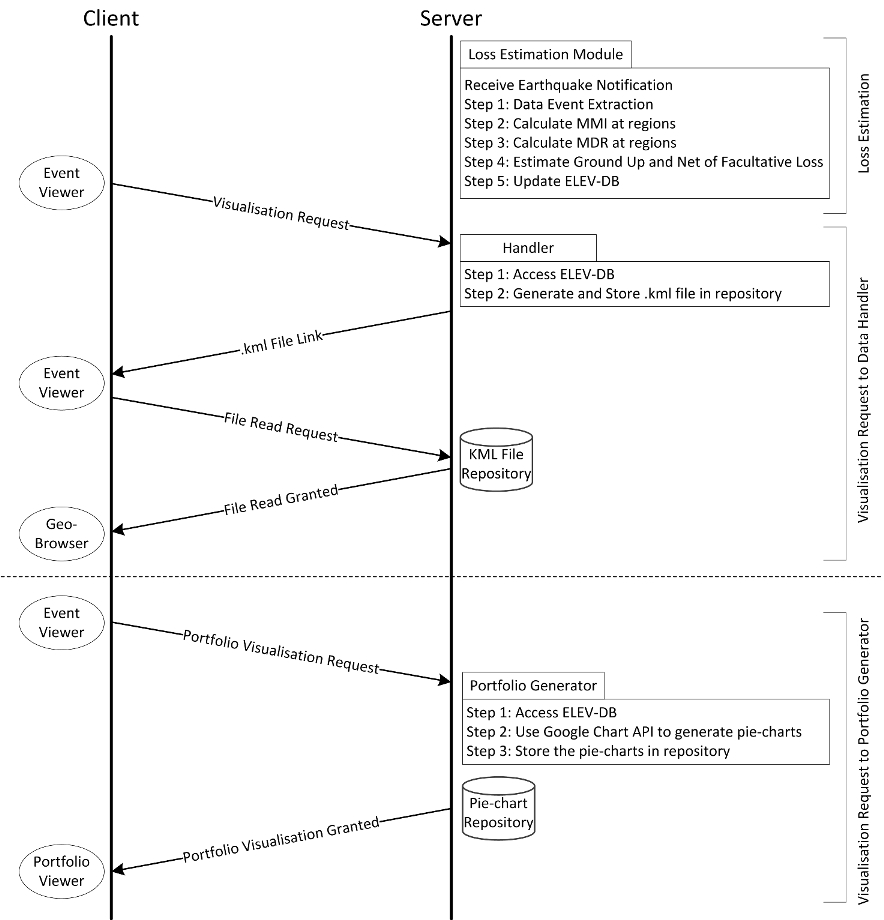}
	\caption{Interaction between client and server system}
	\label{figure4}
\end{figure*}

The client system can raise two type of visualisation requests, those to the data handler and to the portfolio generator. A visualisation request to the data handler is made by the Event Viewer. Based on the type of data that needs to be visualised, the exposure, loss, hazard or static data handlers are invoked. The handler retrieves data from ELEV-DB and a .kml file is generated in the KML File Repository. The Event Viewer after receiving a .kml file link requests to read the file and is accessed by the geo-browser on the client system.

A visualisation request to the portfolio generator again retrieves loss and exposure data from ELEV-DB. The Google chart API is used to generate pie-charts in a repository. The portfolio viewer can then access the pie-charts on the client system. 

\subsection{Benefits of a Distributed Architecture}
There are seven benefits of distributing the modules of APE-ELEV on a server and a client:

\begin{enumerate}
\item[(i)] The server system can facilitate archiving for multiple users. This presents the opportunity for a user to manage his workspace and archive earthquakes of his interest. 
\item[(ii)] The server system is accessible to the client but is concealed from the client. Therefore the installation of third party softwares such as ShakeCast Lite and the Thematic Mapping Engine which are used in the development of APE-ELEV is not required on the client system as they are made available from the server. It needs to be however noted that the installation of a geo-browser is mandatory to view .kml files on the client system. 
\item[(iii)] There is no data management on the client system. Since multiple external data sources including real-time earthquake data, exposure data, geography data and geometry data are ingested by APE-ELEV, user management of these data sources would be cumbersome. In distributed APE-ELEV, data management is carried out at the server. 
\item[(iv)] There are no repositories on the client system. Should a user require to analyse a large number of earthquakes, then the KML file and pie-chart repositories can be large. The client system is granted access to the repositories that are situated on the server.
\item[(v)] The database consisting of voluminous data created by APE-ELEV is resident on the server system. The data is voluminous due to the integration of geometry, geographic, exposure and event data which further produces loss and hazard data at multiple geographic levels. 
\item[(vi)] APE-ELEV can be made globally accessible by hosting the server system on the World Wide Web. 
\item[(vii)] The client system can be made available on multiple platforms such as tablets, smartphones and Personal Digital Assistants (PDAs). The availability of APE-ELEV essentially requires internet access. KML data will require a geo-browser enabled platform.
\end{enumerate}

Administrative privileges to the server will be required for decision makers to be able to use the distributed APE-ELEV to their benefit of not merely interpreting the output of APE-ELEV using the default exposure set but using a custom exposure. As of where the current development of distributed APE-ELEV stands the data management facilitated by the server limits the user ability to adjust input data and customise the output data; the centralised system lends itself more to such custom user requirements. Consequently, multiplier indices considered in Section \ref{validationstudy} cannot be set by the user and this flexibility needs to be incorporated in future research. 

\section{Experimental Studies}
\label{experiments}
This section in the first instance considers the experimental platform and the user interface of APE-ELEV, followed by feasibility and validation studies of the APE-ELEV model. The feasibility of APE-ELEV is confirmed using a test case earthquake of magnitude 9.0 that occurred on 11th March 2011, commonly known as the Tohoku earthquake or referred to as Near the East Coast of Honshu, Japan with an Event ID USC0001XGP in PAGER. The validation study considers 10 global earthquakes and the expected losses computed by APE-ELEV is compared against normalised historic loss data. The validation study is also pursued to determine the probability of the expected losses falling within a pre-defined loss threshold.

\subsection{Experimental Platform}

The data related to the earthquake was available on the PAGER archive \citep{PAGERArchive} and ShakeMap archive \citep{ShakeMapArchive}. The Event Data Extractor in the APE-ELEV architecture fetches data related to the event from the PAGER archive in .xml format and instantiates the ELE module. After the ELE module is instantiated, the losses are estimated as considered in Section \ref{estimator}. The EV module is then employed to visualise the estimated losses.
 
Geometry data for the geographic levels was obtained from the Global Administrative Areas Database \citep{GAAD}, as shapefiles. The shapefiles obtained were large in size containing accurate boundary specification. Since the experiment reported here was a preliminary test, approximate boundary specifications were sufficient, and therefore the shapefile was simplified using the MapShaper tool \citep{MapShaper}.

Figure 5 is a screenshot of the visualiser. The inline map shown on the screenshot represents the ShakeMap representation of the earthquake. The earthquake related data is shown on the right-hand side of the map. The four visualisers of the EV module are listed under Google Earth Visualisation as Static Data, Exposure Data, Hazard Data and Loss Data. The visualisation techniques (choropleth in the screenshot) are available in a drop-down box. The ShakeMap link presents the ShakeMap on the Google Earth application. The Ground Up and Net of Facultative losses computed by the ELE module are displayed under Global Earthquake Loss Model. The Portfolio Loss link presents four pie charts that compares the losses and exposures by line of business such as industrial, personal and commercial.

\begin{figure*}
	\centering
	\includegraphics[width=0.8\textwidth]{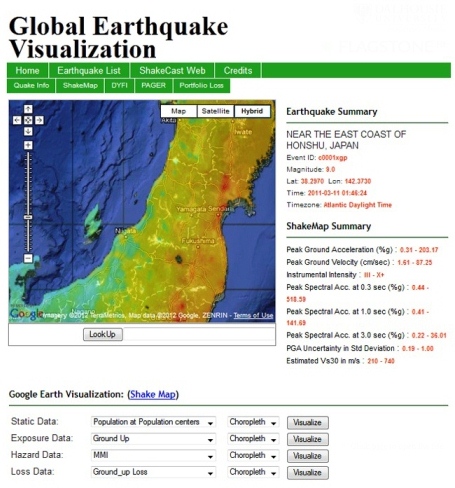}
	\caption{Screenshot of the Visualiser module of APE-ELEV}
	\label{figure5}
\end{figure*}

\subsection{Feasibility Study}

The test case employed in the feasibility study is magnitude 9.0, which occurred in Tohoku, Japan on 11 March 2011 that struck off the Pacific coast of Japan at 05:46 UTC on Friday, 11 March 2011. This recent earthquake was a major catastrophe and affected 28 prefectures. 

It is worthwhile to note that the catastrophe was due to both a tsunami and an earthquake. The APE-ELEV model does not incorporate any mechanism to differentiate between the tsunami and the earthquake related losses. This differentiation, however, is achieved in the model since the input data from USGS PAGER and ShakeCast differentiates the catastrophe by producing earthquake related data. Therefore, the model inherently produces loss estimates for the catastrophe data provided and its accuracy is dependent on the input. 

Figures 6-10 are a set of screenshots obtained from the visualiser. Figure 6 shows the MMI of the affected prefectures using the prism visualisation technique. The gradient scale on the left hand side shows the MMI at the prefectures. The right most pop-up shows GUL and NFL for the earthquake. The pop-up in the centre shows the Exposure, population and hazard data of Shizuoka prefecture. 

\begin{figure*}
	\centering 
	\includegraphics[width=0.8\textwidth]{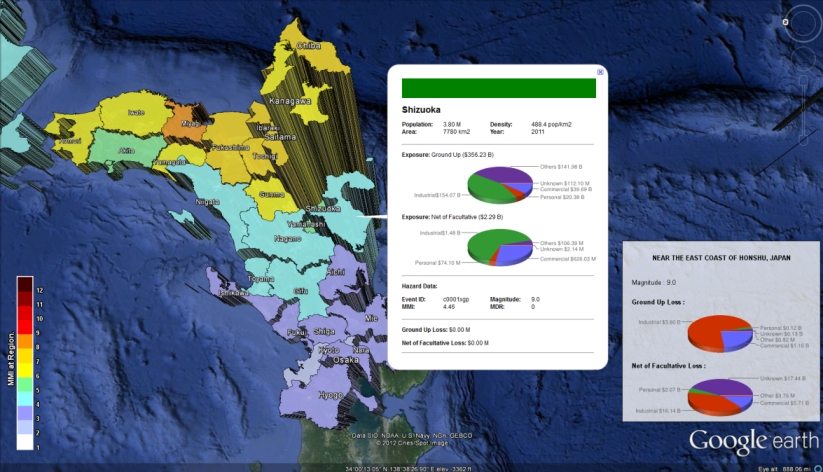}
	\caption{Visualisation of MMI at affected prefectures using prism of experiments from the test-case  Magnitude 9.0, Tohoku, Japan, 11 March 2011} 
	\label{figure6}
\end{figure*}

Figure 7 shows the MDR of the affected prefectures. The choropleth visualisation technique is employed for representing the MDR. The gradient scale on the left hand side shows the MMI at the prefectures. The pop-up shown on the right side shows information relevant to the earthquake for Japan and the pop-up in the centre shows regional information for the Fukushima prefecture. 

\begin{figure*}
	\centering 
	\includegraphics[width=0.8\textwidth]{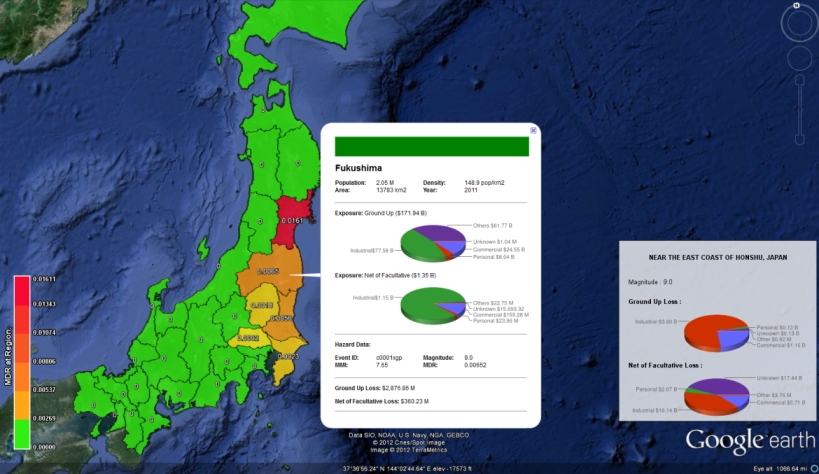}
	\caption{Visualisation of MDR at affected prefectures using choropleth from the test-case  Magnitude 9.0, Tohoku, Japan, 11 March 2011} 
	\label{figure7}
\end{figure*}

Figure 8 shows the superimposition of MDR and population of the affected prefectures. Choropleth is employed for visualising MDR of the prefectures, prisms are employed for visualising NFL and push-pins are used for visualising populations. The two gradient scales on the left side show the scale of MDR and populations. The pop-up shown on the right side shows information relevant to the earthquake and the pop-up in the centre shows regional information relevant to Miyagi prefecture. 

\begin{figure*}
	\centering 
	\includegraphics[width=0.8\textwidth]{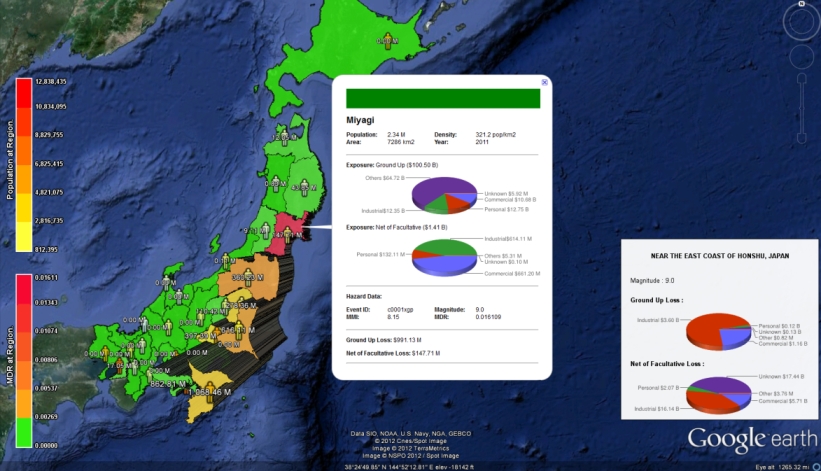}
	\caption{Visualisation of MDR, NFL and population using choropleth, prism and human push-pins respectively from the test-case  Magnitude 9.0, Tohoku, Japan, 11 March 2011} 
	\label{figure8}
\end{figure*}

Figure 9 shows the MMI of the affected prefectures using choropleth, the population in the prefectures using human push-pins and the estimated losses using prisms. The two gradient scales on the left side show the scale of MMI and population. The pop-up on the right side shows the estimated loss information for the entire event in the GUL and NFL categories. The pie charts indicate the losses for industrial, personal, commercial and other lines of business for the exposure data used.

\begin{figure*}
	\centering 
	\includegraphics[width=0.8\textwidth]{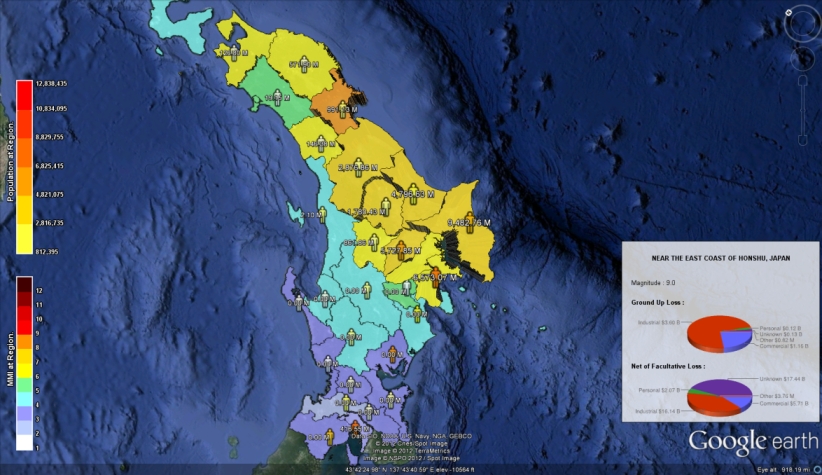}
	\caption{Visualisation of MDR, NFL and population using choropleth, prism and human push-pins respectively from the test-case  Magnitude 9.0, Tohoku, Japan, 11 March 2011} 
	\label{figure9}
\end{figure*}

Figure 10 shows a different view of information visualised in Figure 9. The MMI of the affected prefectures using choropleth, the population in the prefectures using human push-pins and the estimated losses using prisms. MMI and population are shown on the gradient scale. While the right-most pop up showing the pie charts indicates the loss for the entire event, the pop up in the centre shows the losses specific to the Saitama prefecture. The GUL and NFL aggregated for the prefecture along with information relevant to the prefecture and the event are presented.  

\begin{figure*}
	\centering 
	\includegraphics[width=0.8\textwidth]{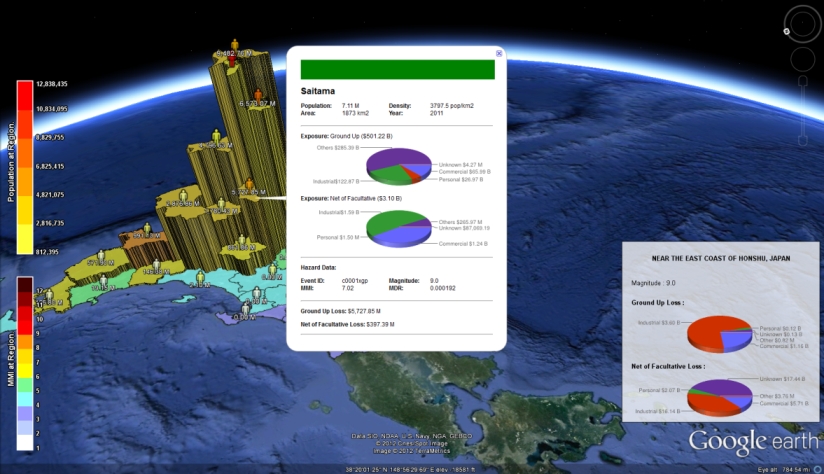}
	\caption{Another view of MDR, NFL and population using choropleth, prism and human push-pins respectively from the test-case  Magnitude 9.0, Tohoku, Japan, 11 March 2011} 
	\label{figure10}
\end{figure*}

Figures 11-18 are screenshots of different alert versions, $A_{1}-A_{15}$ of the test-case earthquake which shows the evolving view of the earthquake and how losses can be rapidly estimated. The MMI of the affected prefectures are shown using choropleth visualisation technique and the height of the prisms are indicative of the Ground Up losses. $A_{1}-A_{5}$ were received within the first day after the event, $A_{6}-A_{8}$ within the same week after the event, $A_{9}-A_{12}$ within the same month after the event and the remaining alerts within six months after the event. 

Figure 11 is based on the first alert, $A_{1}$ which presented data for an overall magnitude of 7.9 twenty two minutes and fifty eight seconds after the event occurred. In this alert, as shown in the figure fourteen prefectures are affected - six prefectures with MMI VII (dark yellow), six prefectures with MMI VI (light yellow) and two prefectures with MMI V (green). The ground up loss for the prefectures are estimated and presented above the prisms indicative of the magnitude of the loss. The estimated losses are highest for the Chiba and Kanagawa prefectures. 

\begin{figure}
	\centering 
	\includegraphics[width=0.5\textwidth]{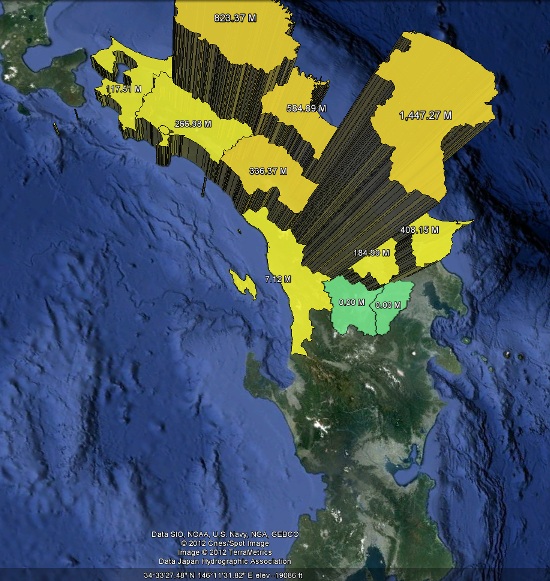}
	\caption{Screenshots of alert version $A_{1}$ of magnitude 9.0, Tohoku, Japan, 11 March 2011 earthquake} 
	\label{figure11}
\end{figure}

Figure 12 is based on the third alert, $A_{3}$ which presented data data for an overall magnitude of 8.8 one hour and fifteen minutes after the event occurred. In this alert, more data was available and was used to update the first alert. While there is a difference in the data showing the magnitude of the earthquake, the MMI data and the estimates for the ground up loss remained the same.    

\begin{figure}
	\centering 
	\includegraphics[width=0.5\textwidth]{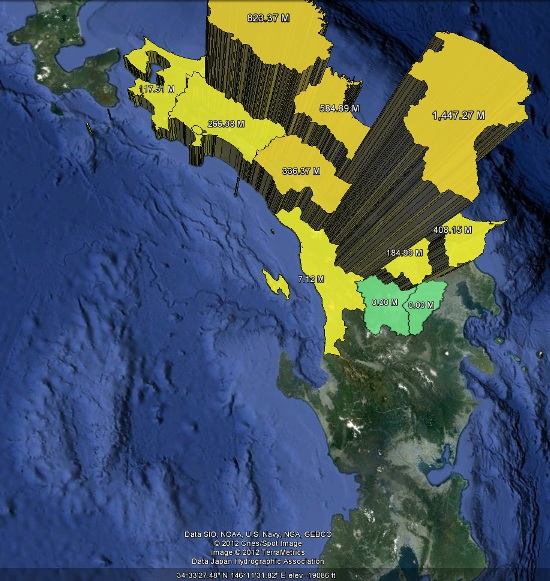}
	\caption{Screenshots of alert version $A_{3}$ of magnitude 9.0, Tohoku, Japan, 11 March 2011 earthquake} 
	\label{figure12}
\end{figure}

Figure 13 is based on the fifth alert, $A_{5}$ which presented data for an overall magnitude of 8.9 two hours and forty four minutes after the event. The MMI information of the prefectures were updated - six prefectures with MMI VII (dark yellow), eight prefectures with MMI VI (light yellow), five prefectures with MMI V (light green) and three prefectures with MMI IV (light blue). The loss estimates for the prefectures have rapidly changed after this alert. For example, for the Chiba and Kanagawa prefectures the ground up loss estimates have increased by approximately 8 times after the first and third alert. The sensor data in this alert has gathered more information about the prefectures which are land-locked. 

\begin{figure}
	\centering 
	\includegraphics[width=0.5\textwidth]{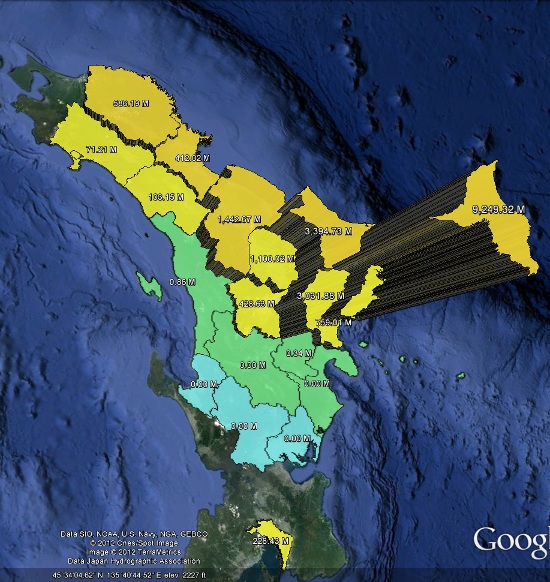}
	\caption{Screenshots of alert version $A_{5}$ of magnitude 9.0, Tohoku, Japan, 11 March 2011 earthquake} 
	\label{figure13}
\end{figure}

Figure 14 is based on the seventh alert, $A_{7}$ which presented data for an overall magnitude of 9.0 four days and nine hours after the event. Again the MMI information of the prefectures are updated with more accurate information gathered by the sensors. One prefecture has an MMI VIII and the ground up loss estimates of the prefectures around Chiba and Kanagawa prefectures have increased. More prefectures to the south of the island have an MMI IV though the losses estimated here are zero.  

\begin{figure}
	\centering 
	\includegraphics[width=0.5\textwidth]{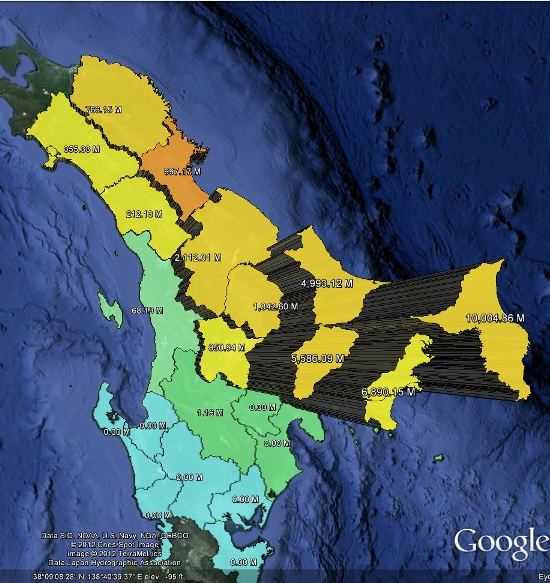}
	\caption{Screenshots of alert version $A_{7}$ of magnitude 9.0, Tohoku, Japan, 11 March 2011 earthquake} 
	\label{figure14}
\end{figure}

Figure 15 is based on the ninth alert, $A_{9}$ which presented data for magnitude similar to the previous alert and was received one week and one day after the event. The data for the next alerts will remain almost similar with minor details updated. While in the previous alerts an evolving view of the hazard, vulnerability and loss were visualised from this alert a constant view is obtained. Again loss estimates in the prefectures to the vicinity of the coastal prefectures are updated. 

\begin{figure}
	\centering 
	\includegraphics[width=0.5\textwidth]{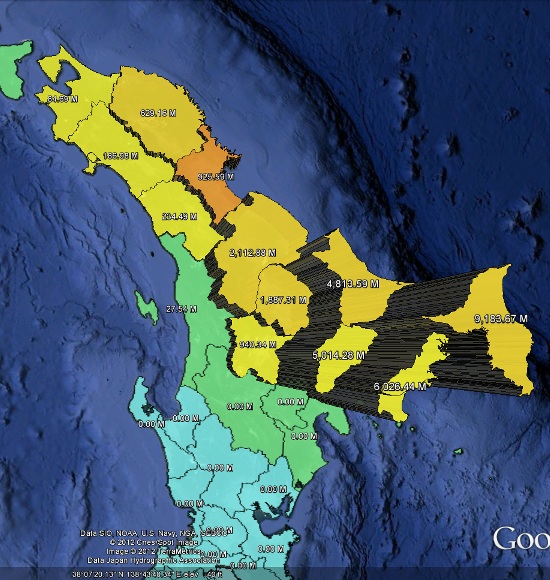}
	\caption{Screenshots of alert version $A_{9}$ of magnitude 9.0, Tohoku, Japan, 11 March 2011 earthquake} 
	\label{figure15}
\end{figure}


Figure 16, Figure 17 and Figure 18 are based on alerts, $A_{11}$, $A_{13}$ and $A_{15}$ respectively. The overall data visualised in these alerts are more or less the same with minimal updates to the MMI and losses estimated for the prefectures. 

\begin{figure}
	\centering 
	\includegraphics[width=0.5\textwidth]{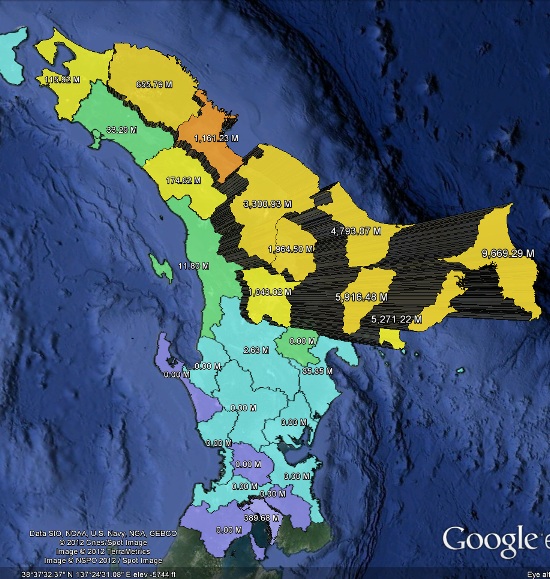}
	\caption{Screenshots of alert version $A_{11}$ of magnitude 9.0, Tohoku, Japan, 11 March 2011 earthquake} 
	\label{figure16}
\end{figure}

\begin{figure}
	\centering 
	\includegraphics[width=0.5\textwidth]{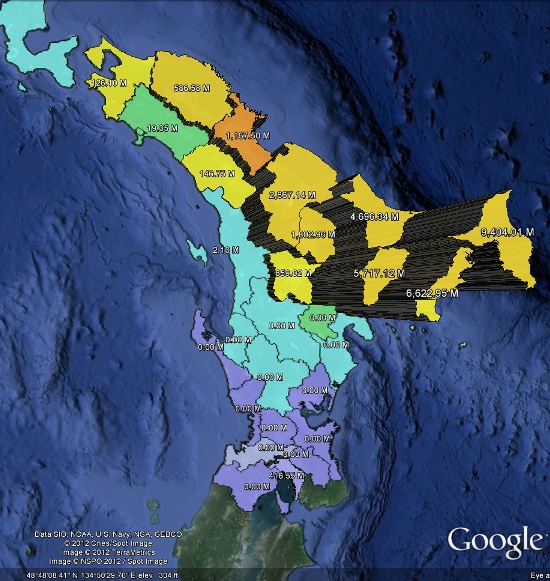}
	\caption{Screenshots of alert version $A_{13}$ of magnitude 9.0, Tohoku, Japan, 11 March 2011 earthquake} 
	\label{figure17}
\end{figure}

\begin{figure}
	\centering 
	\includegraphics[width=0.5\textwidth]{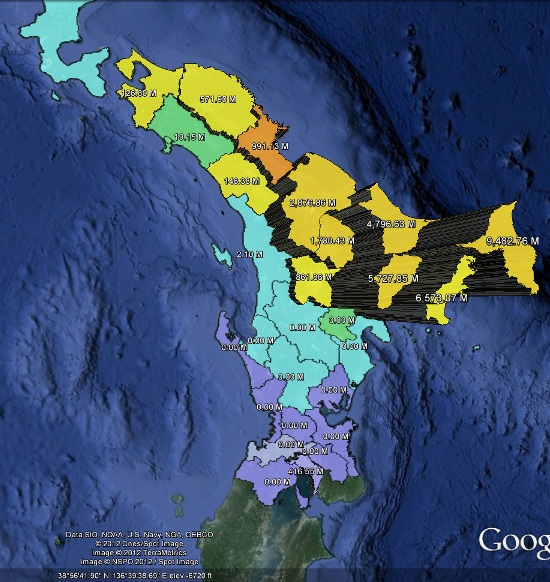}
	\caption{Screenshots of alert version $A_{15}$ of magnitude 9.0, Tohoku, Japan, 11 March 2011 earthquake} 
	\label{figure18}
\end{figure}

\subsection{Validation Study of Loss Model}
\label{validationstudy}

A study that compares the predicted losses of ten global earthquakes against historic loss data was pursued in order to validate the APE-ELEV model. Table 1 shows the list of earthquakes selected for this study, their date of occurrence (dd-mm-yyyy), magnitude, latitude and longitude, historic losses in millions of USD in the year of occurrence of the earthquake, adjustment multipliers to normalise the historic losses to 2012 USD, predicted losses in millions of USD and percent error between the normalised historic and predicted losses. The earthquakes were selected such that (a) they were distributed geographically across different continents, (b) their magnitude was over 5.5, (c) and had occurred in the last 30 years. 

\begin{sidewaystable*}
	\vspace{18cm}
	\begin{center}
		\begin{tabular}{| m{2cm} | m{1.5cm} | c | c | c | c | m{1.4cm} | m{1cm} | m{1cm} | m{1cm} | m{1cm} | m{1.4cm} | m{1.4cm} | m{1cm} |}

			\hline
			\textbf{Region Affected}	&	\textbf{Country}	&	\textbf{Date}	&	\textbf{Mag}	&	\textbf{Lat}	&	\textbf{Long}	&	\textbf{Historic Losses in millions of USD for year y, $D_{y}$}	& \multicolumn{4}{ c |}{\textbf{Adjustment Multipliers}}	&	\textbf{Normalised Historic losses in millions of 2012 USD, $D_{2012}$}	&	\textbf{Predicted Losses in millions of 2012 USD}	&	\textbf{Percent Error \%}\\ \cline{8-11}
				&	&	&	&	&	&	&	\textbf{\tiny{Inflation Multiplier, $IPD_{2012-y}$}}	&	\textbf{\tiny{Inflation-corrected Wealth Multiplier, $ICW_{2012-y}$}}	&	\textbf{\tiny{Wealth Multiplier, $W_{2012-y}$}}	&	\textbf{\tiny{Population Multiplier, $\Delta P_{2012-y}$}}	&	&	&	\\
			\hline \hline
			
			Libertador O'Higgins				&	Chile				& 11/03/2010		& 6.9			&	-34.2592		& -71.9288		&	16.0500		&	1.0558		&	0.9957	&	0.9651	&	1.0318	&	16.8732		&	238.8136	&	1315.33\\ \hline
			WNW of Ferndale						&	USA					& 09/01/2010		& 6.5			&	40.6520			& -124.6920		&	25.0000		&	1.0352		&	0.9850	&	0.9683	&	1.0172	&	25.4904		&	16.8655		&	-33.84\\ \hline
			California							&	USA					& 28/06/1992		& 7.3			&	34.2012			& -116.4360		&	37.8403		&	1.4926		&	1.1539	&	0.9368	&	1.2316	&	65.1718		&	601.4143	&	822.81\\ \hline
			NE of San Simeon					&	USA					& 22/12/2003		& 6.5			&	35.7058			& -121.1010		&	120.7670 	&	1.1981		&	1.0003	&	0.9268	&	1.0793	&	144.7416	&	46.4220		&	-67.93\\ \hline
			\multirow{3}{*}{Sierra El Mayor}	&	USA and Mexico		& \multirow{3}{*}{04/04/2010}		& \multirow{3}{*}{7.2}	&	\multirow{3}{*}{32.2587}	& \multirow{3}{*}{-115.2870}	&	400.0000	&	-			&	-			&	-		&	-		&	421.4479	&	488.3228	&	15.87\\ \cline{2-2} \cline{7-14}
												&	USA					&					&				&					&				&	250.0000	&	1.0302		&	0.9897	&	0.9729	&	1.0172	&	254.9038	&	370.2505	&	-	\\ 
												&	Mexico				&					&				&					&				&	150.0000	&	1.0853		&	1.0230	&	0.9893	&	1.0341	&	166.5441	&	118.0722	&	-	\\ \hline
			California							&	USA					& 18/10/1989		& 6.9			&	37.0400			& -121.8800		&	2,510.0000	&	1.6103		&	1.2119	&	0.9525	&	1.2724	&	4,898.6913	&	7,316.6145	&	49.36\\ \hline
			South Island of New Zealand			&	New Zealand			& 13/06/2011		& 6.0			&	-43.5800		& 172.7400		&	2,816.4549	&	0.9909		&	1.0165	&	1.0099	&	1.0066	&	2,836.7903	&	3,132.1219	&	10.41\\ \hline
			South Island of New Zealand			&	New Zealand			& 21/02/2011		& 6.1			&	-43.6000		& 172.7100		&	13,000.0000	&	1.0025		&	1.0047	&	0.9976	&	1.0070	&	13,093.8628	&	17,660.6445	&	34.88\\ \hline
			California							&	USA					& 17/01/1994		& 6.7			&	34.2130			& -118.5360		&	22,920.0000	&	1.4381		&	1.1106	&	0.9204	&	1.2066	&	36,606.3931	&	4,787.6419	&	-86.92\\ \hline
			Tohuku								&	Japan				& 11/03/2011		& 9.0			&	38.2970			& 142.3730		&	37,200.0000	&	0.9935		&	0.9978	&	0.9873	&	1.0106	&	36,877.4566	&	4,611.4482	&	-87.49\\
			
			\hline	\hline		
 		\end{tabular}
 		\caption{Earthquakes used as test cases in the validation study}
 		\label{Table1}
 		
	\end{center}
\end{sidewaystable*}

The historic data related to all the earthquakes were collected from multiple sources, namely the National Geophysical Data Centre (NSDC) \citep{NSDC}, United States Geological Survey (USGS) \citep{USGS}, PAGER \citep{PAGERArchive}, ShakeMap \citep{ShakeMapArchive}, EM-DAT \citep{EM-DAT} and CAT-DAT \citep{CAT-DAT}. The information collected includes, event data, exposure data, hazard data and loss data. The collected loss data is denoted as $D_{y}$ which are in USD of year $y$ in which the earthquake occurred.  

Normalisation of loss data is reported by \citep{normalisationeg1}, \citep{normalisationeg2}, \citep{normalisationeg3} and \citep{normalisationeg4}. In this paper, the historic loss data is normalised to 2012 USD, denoted as $D_{2012}$ using the normalisation method described by \citep{normalisation1} and \citep{normalisation2}. Three adjustment multipliers are used for the normalisation. Firstly, the Inflation multiplier, denoted as $IPD_{2012-y}$, which uses the implicit price deflator (IPD) for gross domestic product metric sometimes also referred to as GDFDEF. Using this metric any output obtained at the current price is converted into constant-dollar GDP by taking inflation into account. How much change in a base year's GDP is dependent on the changes in the price level is captured by the metric. This metric is available from Economic Research of the Federal Reserve Bank of St. Louis \citep{FRED} and the US Bureau of Economic Analysis \citep{BEA} are employed. 

Secondly, the Population multiplier, denoted as $\Delta P_{2012-y}$, which is the ratio of the population in 2012 and the year of occurrence of the earthquake. The population data is available from the census data published by governmental agencies. 

Thirdly, the Wealth multiplier, denoted as $W_{2012-y}$ is computed as $\frac{ICW_{2012-y}}{\Delta P_{2012-y}}$. $ICW_{2012-y}$ for year, $y$ normalised to 2012 is the Inflation-corrected wealth adjustment obtained as $\frac{\text{Ratio of wealth of 2012 to y}}{\text{Ratio of Consumer Price Index of 2012 to y}}$. The Fixed Asset and Consumer Durable Goods (FACDG) metric in a year is used indicative of the wealth in that given year. The computation of fixed assets capture private and governmental assets and the computation of consumer durable goods take into account non-business goods consumed by households. This metric is obtained from the US Bureau of Economic Analysis (BEA). The sole use of the measure of wealth is not indicative of inflation adjustments and therefore the Consumer Price Index (CPI) is taken into account. Further the wealth multiplier are adjusted for population to a per capita basis. The per capita adjustment is taken into account since increase in wealth is dependent on population and the rate of change of wealth and population are different. 

The normalisation equation is 
\begin{eqnarray}
\label{equation2}
D_{2012} & = & D_{y} \times IPD_{2012-y} \times W_{2012-y} \times \Delta P_{2012-y}
\end{eqnarray}

\noindent or can be restated as
\begin{eqnarray}
\label{equation2a}
D_{2012} & = & D_{y} \times IPD_{2012-y} \times ICW_{2012-y}
\end{eqnarray}

If the Implicit Price Deflator (IPD) index of the GDP is taken into account for computing the Inflation-corrected wealth adjustment instead of the Consumer Price Index (CPI), then the normalisation equation is
\begin{eqnarray}
\label{equation3}
D_{2012} & = & D_{y} \times IPD_{2012-y} \times \frac{ICW_{2012-y}}{\Delta P_{2012-y}} \times \Delta P_{2012-y}\\
		 & = & D_{y} \times \text{Ratio of wealth of 2012 to y}
\end{eqnarray}

In the research reported in this paper, however, $D_{2012}$ is computed using Equation \eqref{equation2} which uses both IPD and CPI. The equation takes into account the effect of population based on the consumption (definition of CPI) in normalisation. However, there is no direct dependence on population as seen in Equation \eqref{equation2a} and Equation \eqref{equation3}. There are challenges in considering the population for earthquake losses. For example, consider an area that was affected by a major earthquake 20 years ago and was sparsely populated then which resulted in minimal ground up loss. For normalising the loss of that earthquake in 2012 factors such as how densely populated that area was in 2012 and the ground up loss if the earthquake occurred in 2012 needs to be considered. For such a consideration regional population statistics will need to be incorporated into the equation. 

Consider for example the earthquake that affected WNW of Ferndale, USA on 9 January 2010 with a magnitude of 6.5. The historic loss for this earthquake in 2010 US dollars is 25 million, represented as $D_{2010}$. The $D_{2010}$ value needs to be normalised for 2012 USD denoted as $D_{2012}$. 
 
The Implicit Price Deflator index in 2010 normalised for 2012, represented as $IPD_{2012-2010}$ can be obtained as the ratio of the Implicit Price Deflator in 2012 ($IPD_{2012}$) to the Implicit Price Deflator in 2010. In 2012, $IPD_{2010}$. $IPD_{2012} = 114.599$ and $IPD_{2010} = 110.702$\textsuperscript{1}\footnotetext[1]{http://research.stlouisfed.org/fred2/data/GDPDEF.txt}. Therefore, $IPD_{2012-2010} = \frac{114.599}{110.702} = 1.0352$. 

Computing the Wealth multiplier index for 2010 normalised to 2012 denoted as $W_{2012-2010}$ requires the computation of two indices, namely the Inflation Corrected Wealth multiplier index ($ICW_{2012-2010}$) and the Population multiplier index ($\Delta P_{2012-2010}$). 

The Wealth of USA in 2012 is 51,117.4 billion USD and the Wealth in 2010 is 48,758.9 billion USD computed from the Fixed Assets and Consumer Durable Goods Account\textsuperscript{2}\footnotetext[2]{http://bea.gov/iTable/iTable.cfm?ReqID=10\&step=1\\\#reqid=10\&step=3\&isuri=1\&1003=16}. Therefore, the Ratio of Wealth of 2012 to 2010 is $\frac{51,117.4}{48,758.9} = 1.0484$. The Consumer Price Index (CPI) for 2012 is 231.227 and for 2010 is 217.230\textsuperscript{3}\footnotetext[3]{http://research.stlouisfed.org/fred2/data/CPIAUCSL.txt}. The Ratio of the Consumer Price Index of 2012 to 2010 is computed as $\frac{231.227}{217.230} = 1.0644$. $ICW_{2012-2010}$ is obtained by dividing the ratio of wealth and the ratio of CPIs of 2012 to 2010, which is $\frac{1.0484}{1.0644} = 0.9850$.

The population of US in 2012 was 314,055,800 and the population in 2010 was 308,745,538. Therefore, the Population multiplier index, $\Delta P_{2012-2010} = \frac{314,055,800}{308,745,538} = 1.0172$. 

The Wealth multiplier index, $W_{2012-2010}$ can then obtained as $\frac{0.9850}{1.0172} = 0.9683$. 

Therefore, for the US earthquake in 2010, normalisation in 2012 US dollars is obtained as 
\begin{eqnarray*}
\label{equation4}
D_{2012} & = & D_{2010} \times IPD_{2012-2010} \times W_{2012-2010} \times \Delta P_{2012-2010}\\
	 	 & = & 25 \text{ million} \times 1.0352 \times 0.9683 \times 1.0172\\
	 	 & = & 25.4904 \text{ million USD}
\end{eqnarray*}

PAGER data (MMI at city level, affected cities due to an earthquake) for global earthquakes are only available after 2007. Therefore, for earthquakes prior to 2008 a in-house computer script was developed to extract data from two sources. The first source was a list of cities whose population is greater than one thousand people. This list is provided by Geonames \citep{Geonames} and contains all the cities in the world whose population is more than one thousand. The model assumes population as point values for cities in all its computations. However, in reality population is a gradient, and the loss estimation technique presented cannot take into account its continuous nature and underestimates the computation of loss taking into account centres with less than a thousand people. The second source was the ShakeMap file which is a representation of the affected grid on a map due to an earthquake and comprises a large set of point data (latitude, longitude and the MMI at that point). The script extracts the list of cities that are affected within the grid and their MMIs. The cities are mapped onto their respective regions using the latitude and longitude information. The exposure data for the geographic levels are collected from publicly available sources. 

The above inputs were used to calculate losses using the method in the APE-ELEV model. As shown in Equation \eqref{equation1}, the MMI at the city level is used to compute the MDR at the same level using the Jaiswal and Wald MDR model, either by direct comparison or by interpolation. The exposure data, which is available for higher geographic levels, is disaggregated onto the city level based on population. The losses for a region are then computed by calculating the sum of the losses for individual cities (loss for individual cities can be computed by the product of the exposure and MDR at the city) within that region. 

A number of obstacles were encountered during the validation study, which are as follows:
\begin{itemize}
\item[(i)] Exposure data had to be collected from a number of disparate sources and was not easy to obtain. 
\item[(ii)] Hazard data is not readily available for events preceding 2008. To collect data for events prior to 2008, as presented above, an in-house script had to be developed. 
\item[(iii)] As data obtained from multiple sources which do not follow a standard convention were integrated in the validation study, significant efforts had to be made towards ordering and organising data and eliminating irrelevant information from the sources. 

\end{itemize}

Despite the above obstacles, (a) event data was easily collected, (b) population data was publicly available and (c) the MMI to MDR was straight forward to calculate based on the vulnerability curves used in PAGER. 

Two column charts were generated based on increasing historic losses. In Figure 19, the predicted and historic losses are shown in millions of USD for events with historic losses less than 1 billion USD, and in Figure 20, for events with historic losses greater than 1 billion USD.

\begin{figure*}
	\centering 
	\includegraphics[width=0.85\textwidth]{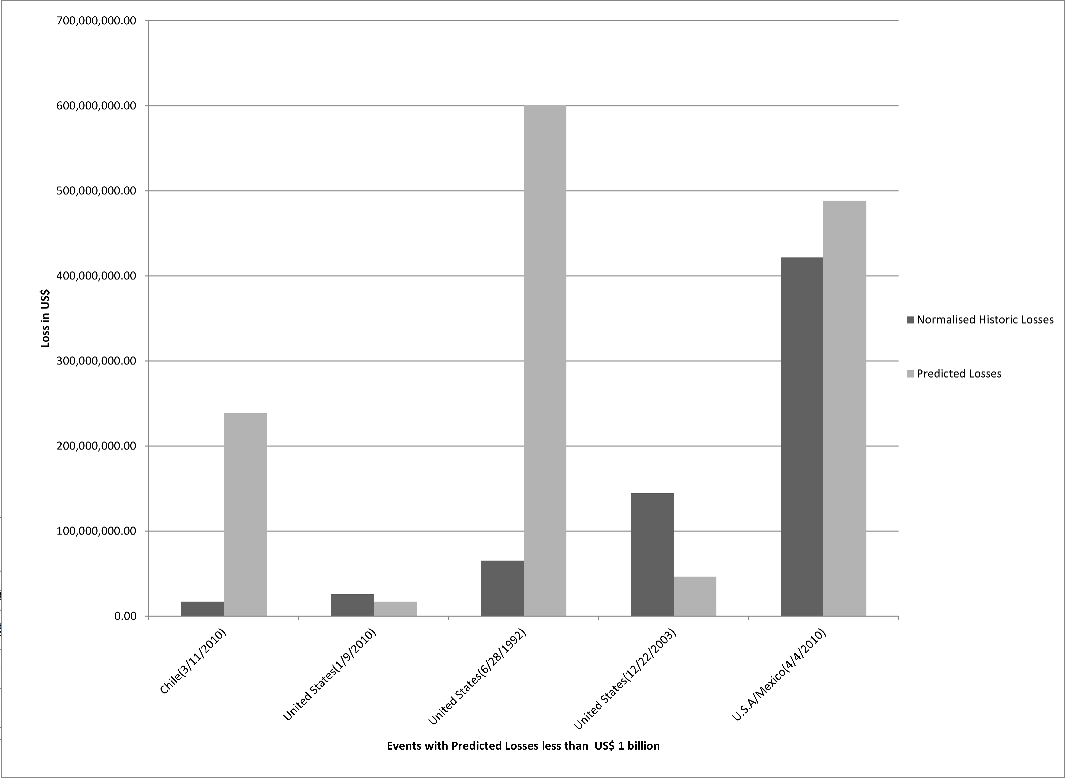}
	\caption{Column charts for historic losses less than 1 billion USD and predicted losses for earthquakes shown in Table 1} 
	\label{figure19}
\end{figure*}

\begin{figure*}
	\centering 
	\includegraphics[width=0.85\textwidth]{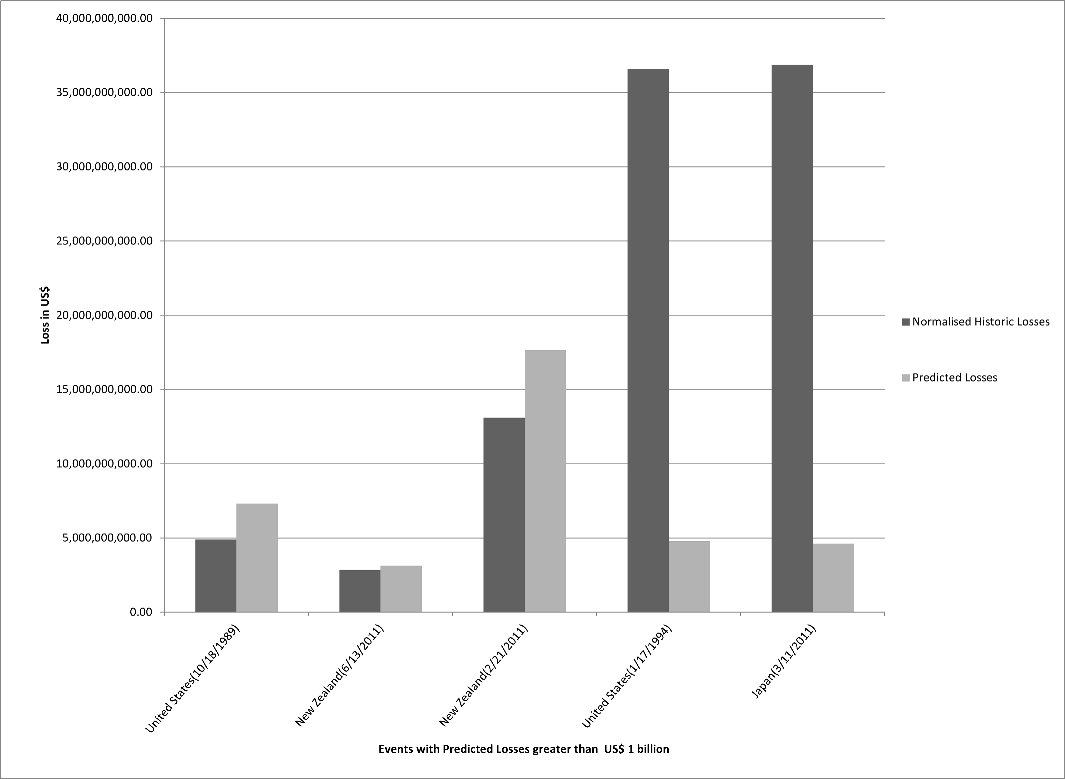}
	\caption{Column charts for historic losses greater than 1 billion USD and predicted losses for earthquakes shown in Table 1} 
	\label{figure20}
\end{figure*}

There are multiple sources of error in the validation study and are as follows: 
\begin{enumerate}
\item[(i)] Input Errors, which refer to the flaws and inaccuracy in the input data to the model. Cities with a population of over 1000 were only considered. This data is constructed on the assumption that population is a discrete distribution, while in reality it is continuous (population outside a city with less than 1000 human inhabitants is not considered). The population data obtained from geonames was inaccurate since a large number of cities presented zero population. This was partially overcome by doing manual look-ups with other reliable sources. However, conflicts with the dates of census of the geonames and the source of the manual look-ups persisted. 

\item[(ii)] Application Errors, which refer to the inaccuracies and assumptions that exist within the model. The MMI of a city was converted to a MDR value using country-based MMI-MDR curves. The assumption here is that every city follows the same curve (values) as of its country. The losses for a few events are calculated in the currency of its country of origin. The value of the currency is then converted to US dollars based on an average conversion rate for the year in which the event occurred.

\item[(iii)] Benchmark Errors, which refer to the assumptions that exist in setting a benchmark. A range of values are available for historic insured losses. It is difficult to determine which value needs to be selected as the benchmark for comparison against the predicted loss. For certain events, historic insured losses were not available, and therefore, the total economic losses were used to estimate the insured loss. This was based on a countrywide take-up rate which may not be accurate for certain regions in a country. 
\end{enumerate}

It is observed that there are two events from the sample which have over 100\% error. The first event affected California in 06/28/1992 with a magnitude of 7.3 , have significant error. This is likely because the most recent exposure for California was only available for the validation study, thereby leading to a significant over-prediction. The second event occurred on 03/11/2010 in Chile with a magnitude of 6.9. The over-prediction is in part likely due to the fact that exposure was disaggregated based on population. In this case, the assumption that exposure is proportional to population is less accurate since only one city with a population of over 1000 was affected. 

The seven events that have less than 100\% error indicate the model is feasible. Further accuracy can be achieved by calibrating the model. 

The loss predicted by the APE-ELEV model is a mean value for an earthquake. To study the probability of a loss threshold $(a,b)$ the $\phi$ distribution which is the standard normal cumulative distribution function is employed as follows \citep{14}:
\begin{equation}
\label{equation9}
P(a < L \leq b) = \phi \left[ \frac{ln(b) - \mu_{ln(L)}}{\zeta}\right] - \phi \left[ \frac{ln(a) - \mu_{ln(L)}}{\zeta}\right]
\end{equation}

\noindent where $\mu_{ln(L)}$ is the predicted value of the logarithm of loss obtained from the model and is assumed to be a lognormal random variable, and $\zeta$ is the normalised standard deviation of the logarithm of loss obtained from \citep{14b}. 

Figure 21 shows the estimate of probability of different loss thresholds ($0 < 1$, $1 < 10$, $10 < 100$, $100 < 1,000$, $1,000 < 10,000$, $10,000 < 100,000$, $100,000 < 1,000,000$) represented in millions of USD for the earthquakes of Table 1. These loss thresholds best represent magnitude losses and are therefore chosen for validating the results in this paper. Different thresholds can be used by appropriately setting $a$ and $b$ values in Equation \eqref{equation9}.

\begin{figure*}
	\centering 
	\includegraphics[width=0.72\textwidth]{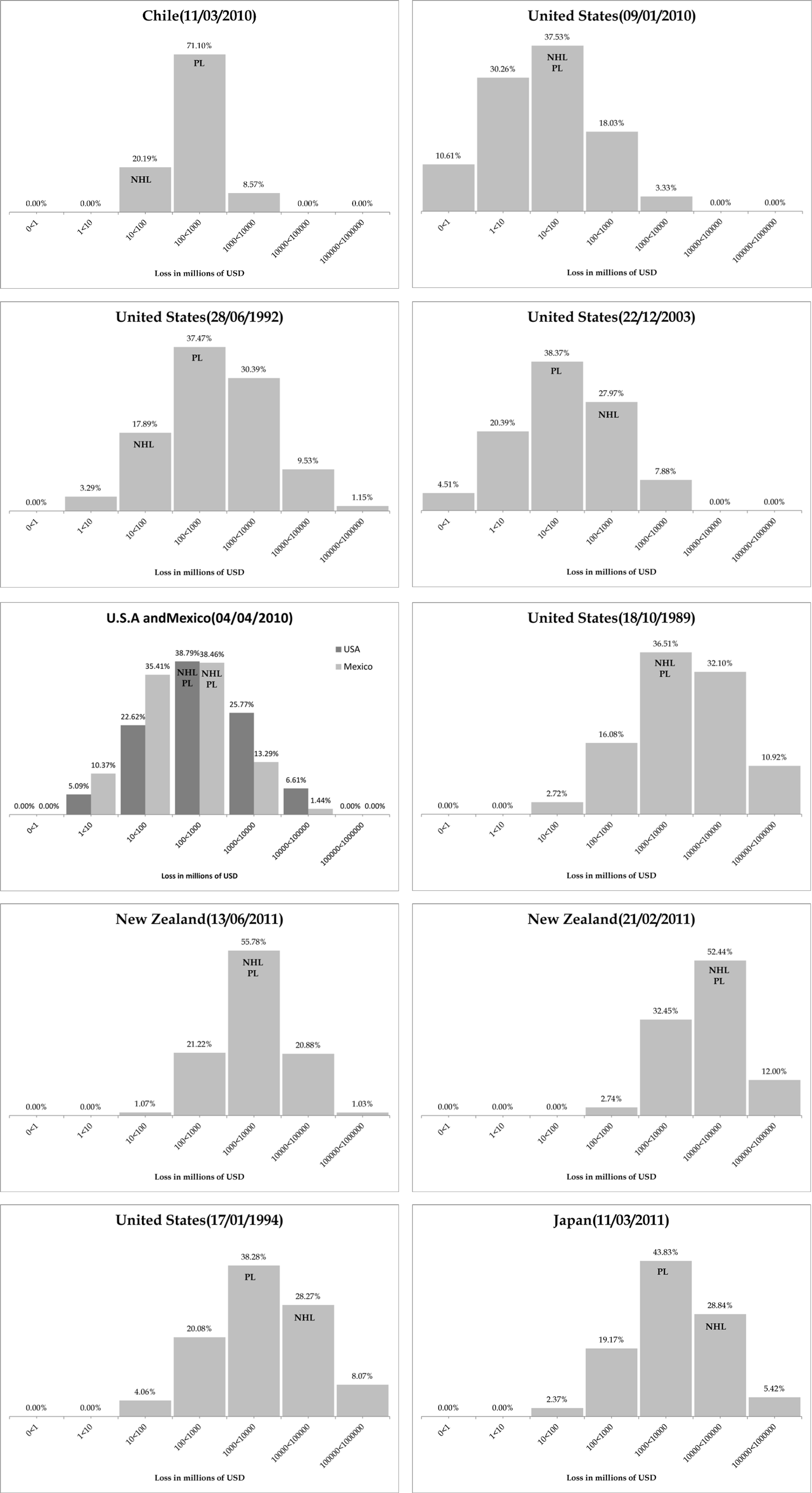}
	\caption{Probability of loss thresholds for the earthquakes in Chile - 11/03/2010, United States - 09/01/2010, United States - 28/06/1992, United States - 22/12/2003, United States and Mexico - 04/04/2010, United States - 18/10/1989, New Zealand - 13/06/2011, New Zealand - 21/02/2011, United States - 17/01/1994, Japan - 11/03/2011} 
	\label{figure21}
\end{figure*}

In this section we have evaluated the performance on APE-ELEV both in terms of how well its data acquisition and visualisation facilities are able to capture the evolving history of earthquake alerts and the performance of its simplistic loss model. The Tohoku earthquake used in evaluating the feasibility demonstrates how data can be rapidly ingested from multiple sources to visualise earthquake alerts as the data related to the event evolves over hours, days and months after its occurrence. 

Evaluation of loss models is tricky at best due to the inherent difficulty in collect consistent exposure and loss data for historic events. In the case of APE-ELEV is important to remember that the goal is to produce on a global basis a crude loss estimate rapidly, as an event evolves, based on very limited information. In this context, the distribution of expected losses is much more important than the point estimates. Our validation demonstrated that the methodology pioneered in PAGER \citep{14} for economic loss can be usefully applied in the context of portfolio losses. 

In 50\% of our evaluation events the observed historical losses and the predicted losses fall into the same loss threshold. In 90\% of our test events the observed historical losses and the predicted losses fall into the two highest loss thresholds. Given the limited data, the loss model gives reasonable order of magnitude estimates, but it is important that users be aware of the inherent limitations of the underlying approach.

\section{Conclusion \& Future Work}
\label{conclusion}
In the time line of an earthquake the sensory data provided by sources such as PAGER/ShakeMap evolves over time. For example, sensory data was updated fifteen times for the Tohoku earthquake ranging from within an hour to six months after the earthquake. The data was first issued twenty three minutes after the earthquake and updated four times during the first day alone. Not only did the earthquake event unfold over time but the data describing the event and our knowledge of the event evolved. The data available initially alone is not sufficient to produce reliable loss estimates. Therefore, analysis of an event soon after it has occurred is challenging and important to generate reliable loss estimates. 

For an earthquake model to be useful in days and weeks after the event, it needs to support (a) rapid data ingestion, (b) rapid loss estimation, (c) rapid visualisation and integration of hazard, exposure, and loss data from multiple sources, and (d) rapid visualisation of hazard, exposure and vulnerability loss data at multiple geographic levels. This paper has presented the design and development of such a model, APE-ELEV (Automated Post-Event Earthquake Loss Estimation and Visualisation). The model comprises three modules, firstly, the Earthquake Loss Estimator (ELE), the Earthquake Visualiser (EV) and the ELEV Database (ELEV-DB). The ELE module is built on relying multiple data sources for accessing real-time earthquake data. Financial losses relevant to the insurance and reinsurance industry are particularly taken into account in the model and are estimated at different geographic levels. The visualisation of the losses on a geo-brower is facilitated by the EV module. The ELEV-DB module aids the cohesive functioning of the ELE and EV modules. 

The recent Tohoku earthquake is used as a test case to demonstrate the feasibility of the APE-ELEV model and how an evolving view of the event is generated using the model. Two types of losses, namely Ground Up and Net of Facultative losses are computed for the earthquake. Further, a set of ten global earthquakes are chosen to validate the model by (a) computing the percentage error between the predicted loss and historic loss values and (b) estimating the probability of loss thresholds for the earthquakes. In the study, all historic loss values are normalised to 2012 US dollars. The key observation is that the model produces reasonable order of magnitude estimates. A video demonstrating a prototype of the distributed APE-ELEV is available at \url{http://www.blessonv.com/software/APE-ELEV}. 

Future work will aim to refine the model by calibrating the PAGER vulnerability curves (for economic losses) for a more accurate use in portfolio insured loss models. A comparison study of estimated losses against normalised historic losses for a larger number of recent earthquake events will be pursued. Extending APE-ELEV for secondary hazards such as tsunamis and floods will be pursued. Efforts will also be made towards augmenting the loss model results with any available historical data points. The distributed APE-ELEV system will be extended for taking custom user input for exposure and catastrophe data and for adjusting the output presentation as required. A study to quantify the input, benchmark and application errors and consider their impact on the estimated loss will be pursued.  

\begin{acknowledgements}
We are grateful to Mr. Philip Shott, Mr. Andrew Siffert and Dr. Georg Hoffman of Flagstone Re's R\&D team, Halifax, Canada for their input and comments. 
\end{acknowledgements}

\pagebreak

\end{document}